# Determining ultrafast carrier dynamics of hybrid perovskites at various stages of nucleation and growth kinetics

Bibek S. Dhami[†], Ravi P.N. Tripathi[†], David J. Hoxie and Kannatassen Appavoo*

Department of Physics, University of Alabama at Birmingham, AL 35294, United States

## Abstract

With hybrid organic-inorganic perovskites increasing its technological reach, from photovoltaics solar cells to light-emitting devices, to nanoscale transistors, it is critical to establish the role of microstructures in dictating how carrier dynamics dictate device efficiency. Here we report on the ultrafast dynamics of charge carriers in hybrid perovskites at various stages of nucleation and growth kinetics. A solution-processed fabrication technique, with spin-coating conditions optimized to control the nucleation density of an intermediate phase, converts to hybrid perovskites upon a temperature gradient annealing. This strategy decouples the nucleation and growth steps that lead eventually to large-grain thin films, allowing us to probe electronic and carrier dynamic differences. We find, surprisingly, that the nucleating microcrystals already display the electronic properties of hybrid perovskites and share similar femtosecond-to-nanosecond dynamics as large-grain hybrid perovskite thin films.

[†]Authors contributed equally
*Address correspondence to appavoo@uab.edu

# Introduction

Hybrid organic-inorganic perovskites (HPs) have shown rapid progress in various energy-conversion devices, especially in photovoltaics, light-emitting diodes and photodetectors. Recently, HPs have also been considered in planar-heterojunction field-effect transistors, and other light-based applications such as optically-pumped nanoscale lasers, low-threshold nonlinear switches[1], phase shifters[2], and electro-optical modulators[3]. Since hybrid perovskites are promising in various optoelectronic devices that operate over a wide range of carrier densities, it is important to understand the impact of carrier density and microstructures on the electronic behavior of hybrid perovskites.

The performance of hybrid perovskites depends critically on their growth conditions. For example, varying the solution-processed methods for thin-film fabrication will dictate grain size, local crystalline heterogeneity, film morphology, and dimensionality (quantum dots, nanowires and bulk crystals). These structural and morphological changes have a profound impact on fundamental processes such as absorption coefficient, photoluminescence efficiency, carrier mobility, optical bandgap, and recombination dynamics[4-6]. In general, the crystallization process in the hybrid perovskites dictate their morphology. Supersaturation of the perovskite precursor is the driving force in the crystallization process, with the perovskite organic and inorganics components that diffuse and reach size greater than the critical radius that initiate the nucleation process. To elicit supersaturation, different approaches have been implemented, like increasing solute concentration, cooling the solution, increasing solvent evaporation, alteration in solvent composition and decreasing solubility of the solute. Other techniques such as hot casting[7-9], anti-solvent treatment[10-12], vacuum quenching[13,14], gas blowing[15], use of additives in precursor solution[16,17], solvent annealing[18], nonstoichiometric composition[19,20], and solvent additives[21] rely on these approaches to control supersaturation and crystallization. In all techniques described, the growth of perovskite crystals involved three processes: the solution reaches supersaturation, nucleation, and subsequent growth towards a large crystal[22]. While the community of hybrid perovskites has intensely investigated how to control the structure and morphology of thin films to improve photovoltaics, little work has relied on controlling nucleating crystals to produce high quality films. In fact, this idea has recently been demonstrated by Sidhik et al., where nucleating seeds of hybrid perovskites were used to create high-quality films that increased device performance. While few reports of in situ structural studies connect the properties of thin films to their parent nucleating crystals, more research is needed to determine how the electronic properties and carrier dynamics of the hybrid perovskites evolve during the growth process.

In this Letter, we determine how the ultrafast electronic properties of prototypical lead halide perovskites evolve during its nucleation and growth kinetic process. We devise a new fabrication strategy to reduce the broad phase space of the solution precursors that can dictate the thin films, allowing us to "freeze" various phases of hybrid perovskites growth on the *same* sample. Such phase-gradient sample enables us to systematically probe the ultrafast broadband optical response of hybrid perovskites at different stages of their growth. To the best of our knowledge, this is the first report of ultrafast spectroscopy conducted on such nucleating microcrystals. Surprisingly, our results show that the microcrystals share very similar optical properties and carrier dynamics with the end product, i.e., the large-grain thin films. By temporally

# Results

Samples of hybrid perovskites (methyl ammonium lead iodide, $CH_3NH_3PbI_3$) were prepared on glass substrates using a modified hot-casting, spin coated solution-processed method. Detailed synthesis process is discussed in the *Methods* section. Briefly, with knowledge that the precursor phases can guide the final hybrid perovskite film morphology[23], we use the spin-coating process to uniformly distribute the nucleating seeds on the substrate. Following this spin-coating step, we employ a temperature-assisted crystallization to yield three different HP phases on the same sample: nucleating microcrystals (dots), needle-like crystals (wires), and large-grain thin films (grains). Each phase can be easily identified using either optical and scanning electron microscopy, and was well separated spatially from each other (Figure 1). Higher and lower temperature regions on the substrate yielded large grains and nucleating microcrystals respectively, while the intermediate zone on the temperature gradient substrate yielded needle-like structures. For the intermediate temperature, the nucleating seeds undergo a one-dimensional growth to form long needles that can grow to hundreds of microns in length within few seconds[24],[25]. Such growth, if allowed to proceed, generates a thin film with stacked-needle morphology and contain many pinholes with low HP coverage. We further characterize each HP phase using X-ray diffraction (XRD), scanning electron and optical microscopy. We found that the dots, wires, and grains had signatures all associated with hybrid perovskite (denoted with an asterisk in figure 1b). A plot of $PbI_2$ is also included in Figure 1b for comparison (grey shaded area). Not surprisingly, for the case of dots and grains, we see the presence of $PbI_2$ signatures ($2\vartheta \sim 12^o$), while unexpectedly, the $PbI_2$ signature is drastically decreased in the wire case. We attribute this effect to the evaporation process of the solute and needle-like growth dynamics that creates highly crystalline material, albeit with poor substrate coverage. At higher substrate temperature, typically $> 125\ ^oC$, spherulitic growth takes place radially from the nucleating seeds that leads to formation of large grains ($10$ -$100\ \mu m$). This growth pattern is confined to a two-dimensional plane, results in films of thickness of $\sim 1\ \mu m$, and terminates within few seconds. The growth stops when two domains meet, creating a grain boundary, with a film morphology defined by a Voronoi diagram[26].

While it is not feasible to probe ultrafast electronic properties of hybrid perovskites during the crystallization process, i.e., while spin-coating, our fabrication strategy enables us to probe various stages of hybrid perovskites growth in an indirect way. To assess the intrinsic carrier dynamics of the various microstructures of hybrid perovskites, we conduct broadband near-infrared femtosecond pump-probe spectroscopy. Briefly, a high repetition rate diode-pumped $Yb{:}KGW$ femtosecond laser system based on chirped-pulse amplification (PHAROS from Light Conversion) is used to pump a hybrid optical parametric amplifier to produce light pulses centered at $840\ nm$ that are subsequently frequency doubled to 420 nm (pump). Changes in transmission of a spectrally dispersed broadband probe pulse beam is recorded when the sample is excited (pump on) and when it is in its ground state (pump off) at varying pump-probe delays and at a repetition rate of $1\ kHz$. Further details of our ultrafast pump-probe setup is found in the *Methods* section. We conducted ultrafast transient spectroscopy in the fluence range between $7.5 - 65\ \mu J/cm^2$. Representative datasets of these broadband ultrafast transient absorption measurements at a pump fluence of $\sim 7.5\,\mu\text{J}/\text{cm}^2$ are shown in Figure 2. The main signal ($\lambda \approx 750\ nm$) corresponds to an electromagnetic induced transparency (bleach) near the bandedge of the hybrid perovskite. Such ground state bleach is caused by the transition between valence-band maxima and conduction-band minima at the R point[27]. Specifically, for each phase, we note only a slight variation in differential absorption minima for the case of grains ($749\ nm$), wires ($746\ nm$) and dots ($756\ nm$). Normalized temporal slices for $\lambda \approx 755 \pm 5\ nm$ and spectral slices at $\tau = 5\ ps$ are shown for the dots, wires and grain phases (Figures 2b and 2c). While there are visible differences in their kinetics, the degree of similarity especially for the differential spectra between the nucleating microcrystals and thin film grains, is remarkable. The differences in kinetics will be further investigated in the following paragraphs.

To quantify the carrier dynamics most pertinent to our system, i.e., near the bandedge dynamics, we examine the population term within the $\chi^{(3)}$ formalism [28,29]. We perform fits on all transient data acquired on each phase at the relevant spectral slice ($\lambda = 750 \pm 5\ nm$) with a fitting procedure that corresponds to a convolution of the system response with the cross-correlation of the pump and probe pulses; this fitting equation to probe the change in carrier density $n(t)$ in halide perovskites is given by [30,31]:

$$\Delta\alpha\ (\lambda, \tau) \propto \frac{dn(t)}{dt} = \varphi(t) - k_1 n(t) - k_2 n(t)^2 - k_3 n(t)^3$$

where 
$$\varphi(t) = \frac{1}{h\nu_{pump}} (1 - e^{-\alpha d}) \left(\frac{1}{d}\right) (1 - \mathcal{R}) \left(\frac{\mathcal{F}}{\tau_{pulse}}\right) e^{\frac{-2t^2}{\tau_{pulse}^2}}$$

Here, $\varphi(t)$ is the carrier generation process. The rate coefficients $k_{i=1,2,3}$ corresponds to first-order (monomolecular) recombination caused trap-assisted processes, second-order (bimolecular) caused by intrinsic electron-hole pair recombination, and third-order (Auger) which is a many-body recombination process of an electron-hole pair that transfer its energy and momentum to a third participant, i.e., either a hole or an electron. The process involves a phonon absorption or emission that assist this recombination process. Other terms in this equation are: $h$ is the Planck's constant, $\nu_{pump}$ is the pump pulse frequency, $\alpha$ is the absorption coefficient, $d$ is the thickness of the thin film, $\mathcal{R}$ is the reflectance of the thin film, $\mathcal{F}$ is the fluence per pulse, $\tau_{pulse}$ is pulse duration as measured by autocorrelation measurement. For such fitting, we assume the ideal case that $k_{i=1,2,3}$ are constants and that the densities of both electrons and holes are similar. Representative fits of the kinetics are overlaid on the experimental data of Figure 2b, and highlights the excellent quality of the fits. With pump excitation fluences in the $7.5 - 65\ \mu J/cm^2$ range, corresponding to $(1\ 10) \times 10^{17}\ carriers/cm^3$, bimolecular recombination ($k_2$) processes is well known to dominate the kinetics. Using the total recombination rate $r(n, t) = k_1 + nk_2 + n^2 k_3$, we find that indeed the bimolecular recombination rate contribute to $> 90\ \%$ of the kinetic trace. For the large grain thin film, we retrieve values for $k_2$ in the range of $(15\ 60) \times 10^{-10}\ cm^3 s^{-1}$. These fluence-dependent values are within typical range, as measured by various spectroscopic techniques. A similar fitting procedure is conducted for the dot and wire phases, and for an excitation fluence $\sim 26\ \mu J/cm^2$, we obtain $k_2$ recombination rates about an order of magnitude larger, i.e., $290 \times 10^{-10}\ cm^3 s^{-1}$ and $470 \times 10^{-10}\ cm^3 s^{-1}$, respectively.

## Discussions

The pseudo-color representation of transient absorption spectra as a function of probe wavelength and pump-probe delay for the various phases highlight that within the first second of nucleation and crystal growth, the electronic and optical properties of hybrid perovskites are established. In all three phases, we observe a pronounced bleaching feature near the optical bandgap of ~$750\ nm$ ($E_g$~$1.65\ eV$). The positive $\Delta\alpha$ signatures ($\lambda < 725\ nm$) correspond to photoinduced intraband absorption of the excited charge carriers. The other positive $\Delta\alpha$ signature at longer wavelengths (i.e., lower energy side of the bandgap), is also visible for all three phases and attest to the electronic properties of the hybrid perovskites. These signatures, that last for less than a picosecond, are assigned to hot carrier cooling dynamics and correlate with the buildup of ground state bleaching. This feature arises from renormalization of the bandgap caused by the presence of free charge carriers. The presence of this signature, again, attest to the electronic properties of the hybrid perovskites, even in the nucleating microcrystal phase.

The carrier dynamics of hybrid perovskites are mostly dominated by bimolecular recombination at all fluence studied. To provide insights on all recombination processes that may take place, like trapping ($k_1$) and Auger ($k_3$), we show plots of (i) $1/(\Delta Abs)^2$, (ii) $1/\Delta Abs$, and (iii) $-\ln(\Delta Abs)$ vs. time delays. These plots provide straight lines in the case of *pure* (i) Auger, (ii) bimolecular, and (iii) trapping behavior within the time frame most relevant to those mechanisms (Figure 3 and 4). From the slopes, we can obtain more physical insights into our sample. For example, to evaluate trapping mechanism in grains (Figure 3, bottom), we examine longer timescale signatures ($\tau > 1\ ns$). For all pump excitation fluence studied, as expected, we find slopes that are independent of excitation density, and $k_1$ is calculated to be ~ $4 \times 10^8\ s^{-1}$. Thus, doubling the excitation density doubles the recombination rate for this mechanism. Once trap-assisted recombination channels have been eliminated, bimolecular recombination sets in. We find bimolecular rates $(14\ 25) \times 10^{-10}\ cm^3 s^{-1}$ using this technique. These values agree with previous experiments $0.2 - 20 \times 10^{-10}\ cm^3 s^{-1}$, which reflect the intrinsic nature of bimolecular recombination for prototypical $CH_3NH_3PbI_3$. In Figure 3 (top), we attempt to quantify Auger contribution on short timescale ($\tau < 100\ ps$), which demonstrate that such mechanism plays a less critical role for the fluences studied. This observation is in good agreement with our global fit where bimolecular recombination was found to dominate.

We now turn to the carrier dynamics of hybrid perovskites for the three phases. Plotting $-\ln(\Delta Abs)$ vs. time delay in Figure 4 (bottom) readily highlights the different slopes at longer timescale, with the grain data displaying a lower monomolecular-trapping rate ($2 \times 10^8\ s^{-1}$) as compared to the wire ($4.5 \times 10^8\ s^{-1}$) and dot ($4 \times 10^8\ s^{-1}$) phases. The values agree with our understanding that the wire and dot phases have more trap-related charge-carrier recombination since their growth might have been quenched, caused by a less than optimum temperature. To study higher-order recombination processes, we also plot in Figure 4 the $1/\Delta Abs$ and $1/(\Delta Abs)^2$ transient data that corresponds to the bimolecular and Auger recombination, respectively. Since it is not strictly feasible to distinguish between second- and third-order, we emphasize that our plots only provide a guide to understanding how dynamics in various microstructured phases of hybrid perovskites are modified. For example, at a fluence of $25\ \mu J/cm^2$, there is about an order of magnitude increase in the bimolecular recombination when we compare grain ($k_2 = 45 \times 10^{-10}\ cm^3 s^{-1}$) to both dot or wire phases ($k_2 = 250 - 400 \times 10^{-10}\ cm^3 s^{-1}$). This increase in rate is attributed to the microstructure that can enhance carrier-carrier recombination. The fact that the wire phase has a more confined geometry than the dot phase and thus has a higher $k_2$ value validates this argument. Moreover, in terms of trap-related recombination rate, both the wire and dot phases have similar $k_1$ values. For timescales that are relevant to Auger recombination, i.e., $\tau < 50\ ps$, we again see an increase of Auger recombination rate of about an order in magnitude. While we emphasize that for the fluence studied here, bimolecular recombination dominates, increase in $k_3$ rates could potentially be related to size effects that increase carrier-carrier interaction to increase recombination rate. Indeed, a change in structure corresponds to a change in the momentum mismatch requirement

# Methods

**Sample Fabrication**

Thin films of hybrid perovskites are synthesized via a temperature-assisted crystallization process. In a typical preparation method, synthesis of HPs thin films were performed in two steps. The first step involves preparing the substrate and the second step comprises the growth of the thin films on preprocessed substrate. The entire process was performed in a nitrogen-filled glovebox. In a typical preparation method, $0.2\ g$ methyl ammonium iodide ($CH_3NH_3I$, Sigma Aldrich, $98\%$) and $0.578\ g$ lead iodide ($PbI_2$, Sigma Aldrich, $99\%$) were mixed with $1.0\ ml$ of anhydrous N,N-dimethylformamide ($DMF$, Sigma Aldrich, $99\%$) in glass vial. The mixture solution was then kept for heating at $110^0$ C, followed by ultrasonication. This ensures the homogeneous mixing of all the constituent compounds in DMF. Next, we processed the microscopic glass coverslip ($2.5\ cm\ \times\ 2.5\ cm$, Fisherbrand®). Glass coverslips were cleaned with isopropanol (Sigma Aldrich, 99%), followed by ultrasonication. Cleaned glass coverslips were then coated with $ZnO$ nanospheres solution (size $< 130\ nm$, $40\ wt\ \%$ in ethanol, Sigma Aldrich). The cleaned substrates were then baked at $100^oC$ to evaporate the ethanol residue. Moreover, ZnO coating ensures uniform heating throughout the glass coverslips. Now, the prepared solution of hybrid perovskites was spin coated on the substrate at $4000\ rpm$ for $5\ seconds$ and transferred back to the heater. To introduce the temperature gradient, the substrate with the precursor solutions was held at an angle of $45^o$, which enables us to obtain three growth phases that are spatially distinct and can be probed optically.

**Ultrafast Spectroscopy**

Time-resolved absorption data of our samples were obtained using transient femtosecond pump-probe spectroscopy. These spectra are acquired by exciting samples with 340 nm pulse from optical parametric amplifier having pulse duration of $40\ fs$, beam spot size of $150\ \mu m$ at two pump fluences relevant to our lasing experiment, i.e., at $30\ \mu\frac{J}{2}$ and at 150 μJ/cm². We conduct near-ultraviolet femtosecond pump-probe spectroscopy using a high repetition rate diode-pumped $Yb{:}KGW$ femtosecond laser system based on chirped-pulse amplification (PHAROS from Light Conversion) to pump a hybrid optical parametric amplifier and produce light pulses centered at 680 nm that are subsequently frequency doubled to 340 nm (pump). The pulses are hereafter frequency doubled using a barium-beta-borate crystal (BBO: $Type\ I$, $\theta = 29.2^0$ and $\phi = 90.0^0$, 0.5 mm thick) to produces pulses at 340 nm (i.e., pump) and focused on the sample of interest with a spot size of $150\ \mu m$ in diameter ($1/e^2$). As probe, we focus $7\ \mu J/pulse$ from our amplifier ($\lambda = 1030\ nm$) on a continuously translated calcium fluoride crystal to generate a vertically polarized white-light supercontinuum. Using a pair of spherical parabolic mirrors, the supercontinuum probe pulses are overlapped spatially with the pump beam on the sample (probe beam diameter of $35\ \mu m$), and synchronized temporally using a mechanical stage. To vary the time delay between probe and pump pulse, the pulse generated by OPA is sent through optical delay line consisting of retroreflector mounted on high precision motorized translational stage. The 680 nm light from the laser is split by beam splitter into probe and pump. The pump pulse having wavelength of 340 nm is generated from BBO crystal and is passed through the chopper and is focused on the sample. The probe pulse which is delayed by using translational stage is passed through the sapphire crystal to get white light continuum. The intensity of the probe is too weak that noany appreciable population will be excited from ground to excited states or vice versa. It is then focused on the sample spatially overlapped with pump, collimated and is sent to detector.

# Figures

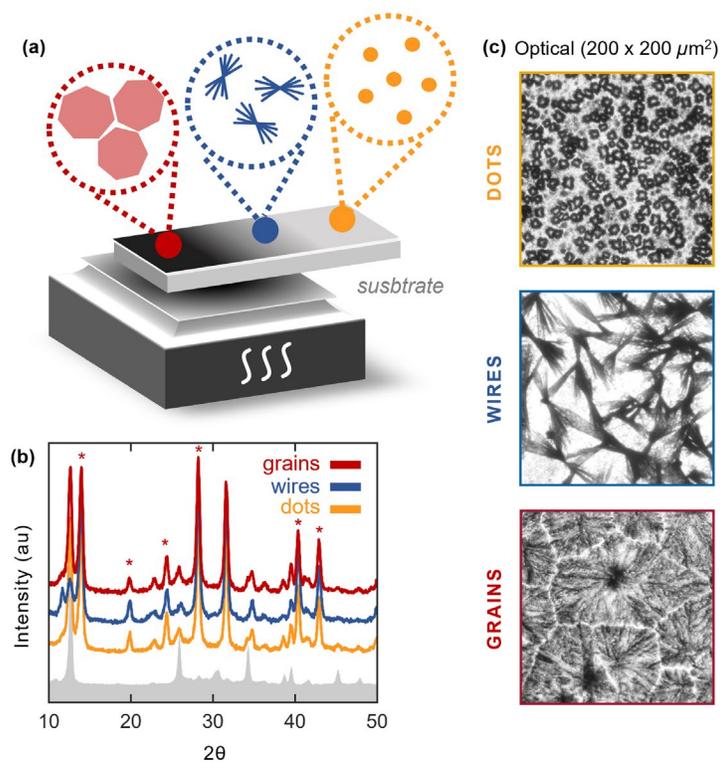

**Figure 1. Fabrication and characterization of hybrid perovskites.**

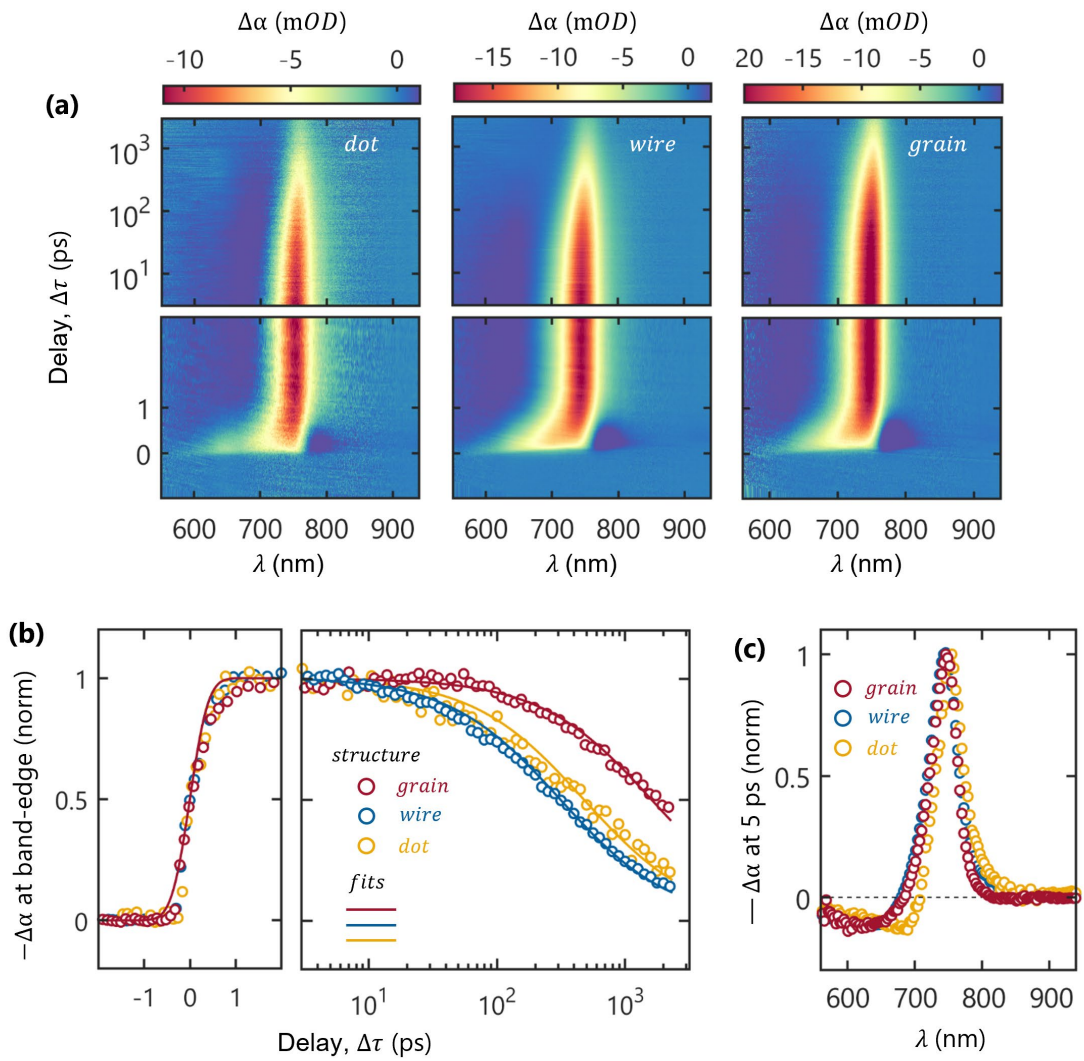

**Figure 2: Ultrafast broadband spectroscopy of hybrid perovskites.**

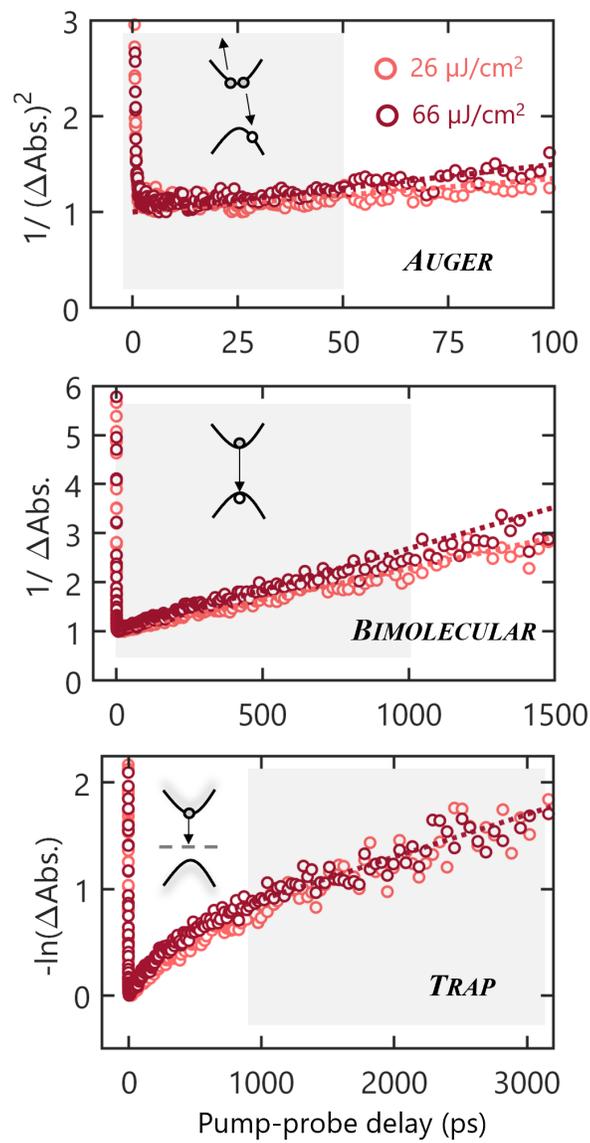

**Figure 3: Fluence-dependent analysis of hybrid perovskite for grain.**

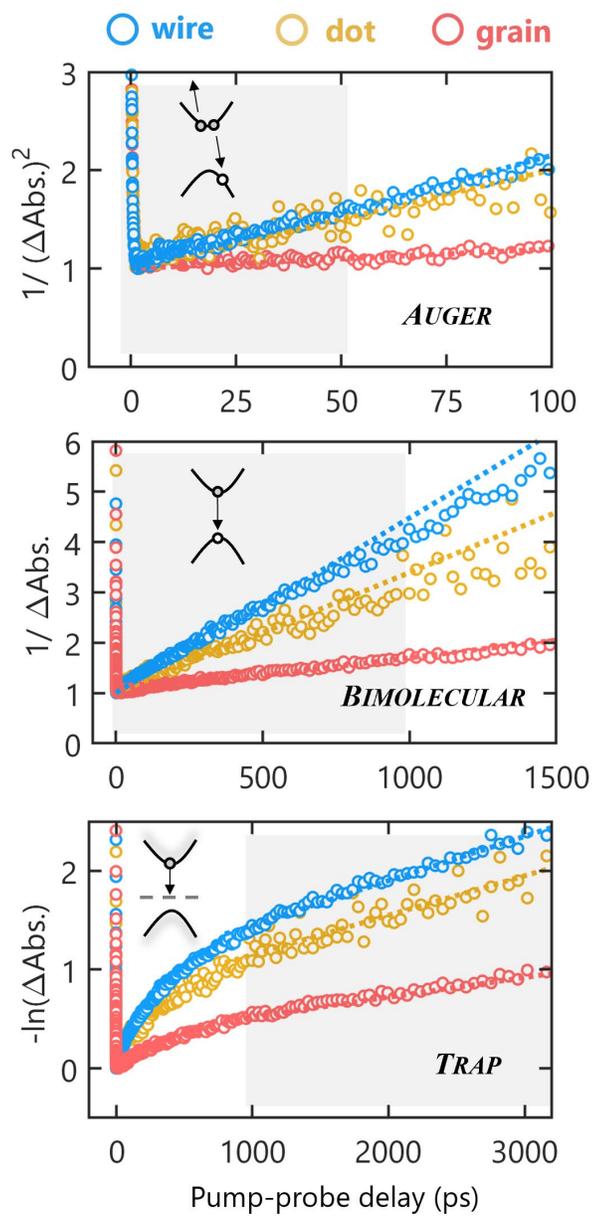

**Figure 4: Structure-dependent analysis of hybrid perovskites.**